\documentclass{article}
\usepackage{spconf}
\usepackage[english]{babel}

\usepackage{amsmath, amssymb, amsthm,commath,mathtools}
\usepackage{tikz}
\usetikzlibrary{calc}
\usetikzlibrary{shapes,arrows,fit,patterns,positioning,decorations.markings}
\usepackage{graphicx}
\usepackage{caption}
\usepackage{dsfont}
\usepackage[caption=false]{subfig}
\usepackage[ruled,vlined,linesnumbered]{algorithm2e}

\let\oldnl\nl
\newcommand{\nonl}{\renewcommand{\nl}{\let\nl\oldnl}}

\newtheorem{proposition}{\bf Proposition}

\newcommand{\DV}[0]{\textit{DV}}
\newcommand{\NWJ}[0]{\textit{NWJ}}

\newcommand{\KL}[2]{\text{D}_{\text{KL}}\!\left(#1\;\lvert\rvert\;#2\right)}

\title{Conditional Mutual Information Neural Estimator}

\name{Sina~Molavipour, Germ\'{a}n~Bassi, Mikael~Skoglund\thanks{This work was supported in part by the Knut and Alice Wallenberg Foundation and the Swedish Foundation for Strategic Research.}}
\address{School of Electrical Engineering and Computer Science\\ KTH Royal Institute of Technology, Stockholm, Sweden\\
	{\tt\{sinmo,germanb,skoglund\}@kth.se}}

\begin{document}
	\maketitle
	
	\begin{abstract}
		Several recent works in communication systems have proposed to leverage the power of neural networks in the design of encoders and decoders.
		In this approach, these blocks can be tailored to maximize the transmission rate based on aggregated samples from the channel.
		Motivated by the fact that, in many communication schemes, the achievable transmission rate is determined by a conditional mutual information term, this paper focuses on neural-based estimators for this information-theoretic quantity.
		Our results are based on variational bounds for the KL-divergence and, in contrast to some previous works, we provide a mathematically rigorous lower bound.
		However, additional challenges with respect to the unconditional mutual information emerge due to the presence of a conditional density function which we address here. 
	\end{abstract}
	%

	\section{Introduction}\label{sec:intro}
	
	Although originally conceived to address communication problems, information-theoretic measures have provided insight in many fields including statistics, signal processing, and even neuroscience.
	Entropy, KL-divergence, and mutual information (MI) are extensively used to explain the behavior and relation among random variables.
	For example, MI and its extensions are used to characterize the capacity of communication channels~\cite{el2011network} as well as to define notions of causality~\cite{quinn2011estimating, Mol2017test}.
	Information-theoretic quantities have also made their way into machine learning and deep neural networks~\cite{tishby1999information}.
	They have been adopted to regularize optimizations in neural networks~\cite{hjelm2018learning} and Markov decision processes~\cite{tanaka2017finite}, or to express the flow of information in layers of a deep network~\cite{gabrie2018entropy}. In a generative adversarial network (GAN), in which a neural network is optimized to perform towards the best-trained adversary, the relative entropy plays an eminent role~\cite{goodfellow2014generative, nowozin2016f}.
	
	On the other hand, neural networks have also been applied in communication setups as part of the encoder/decoder blocks~\cite{o2017introduction, dorner2017deep, fritschek2019deepWTC}.
	Learning the end-to-end communication system is challenging though, and requires the knowledge of the channel model, which it is not available in many applications.
	One solution to this problem is to train a GAN model to mimic the channel, which can later be used to learn the whole system~\cite{ye2018channel, o2018physical}.
	Alternatively, one can design encoders and optimize them to achieve the maximum rate (capacity), which is characterized by the MI.
	In~\cite{fritschek2019deep}, the authors exploit a neural network estimator of the mutual information proposed in~\cite{belghazi2018mine} to optimize their encoders.
	Although estimating the MI has been studied extensively (see e.g., \cite{wang2009universal, paninski2003estimation}), the capacity in many communication setups (such as the relay channel, channels with random state, wiretap channels) is described by the conditional mutual information (CMI), which requires specialized estimators.
	Extending the existing estimators of MI for this purpose is not trivial and is the main focus of this paper. Motivated by~\cite{fritschek2019deep}, we investigate estimators using artificial neural networks. 
	
	The main challenges in estimating the CMI stem from empirically computing the conditional density function and from the curse of dimensionality.
	The latter is, in fact, a common problem in many data-driven estimators for MI and CMI.
	A conventional estimator in the literature for MI is based on the $k$ nearest neighbors ($k$-NN) method~\cite{kraskov2004estimating}, which has been extensively studied~\cite{gao2018demystifying} and extended to estimate CMI~\cite{runge2018conditional, vejmelka2008inferring, frenzel2007partial}.
	In a recent work, the authors of~\cite{belghazi2018mine} propose a new approach to estimate the MI using a neural network, which is based on the Donsker--Varadhan representation of the KL-divergence~\cite{donsker1975asymptotic}.
	Improvements are shown in the performance of estimating the MI between high-dimensional variables compared to the $k$-NN method.
	Several other works also take advantage of variational bounds (\cite{barber2003algorithm}) to estimate the MI and show similar improvements~\cite{oord2018representation, poole2018variational}.
	For estimating the CMI, the authors of~\cite{mukherjee2019ccmi} introduce a neural network classifier to categorize the density of the input into two classes (joint and product); then they show that by training such a classifier, they can optimize a variational bound for the relative entropy and correspondingly for the MI and the CMI.
	Furthermore, they suggest the use of a GAN model, a variational autoencoder, and $k$-NN to generate or select samples with the appropriate conditional density. 
	
	Although the trained classifier proposed in~\cite{mukherjee2019ccmi} asymptotically converges to the optimal function for the variational bound, a relatively high variance can be observed in the final output.
	Consequently, the final estimation is an average over several Monte Carlo trials.
	However, the variational bound used in~\cite{mukherjee2019ccmi} contains non-linearities which results in a biased estimation when taking a Monte Carlo average.
	A similar problem is pointed out in~\cite{poole2018variational} where the authors suggest a looser but linear variational bound.
	In this paper, we take advantage of the classifier technique applied to a linearized variational bound to avoid the bias problem.
	The $k$-NN method is used to collect samples from the conditional density.
	In Section~\ref{sec:pre}, the variational bounds are explained and we shed light on the Monte Carlo bias problem.
	Afterwards, the proposed technique to train the classifier and estimate the CMI is stated in Section~\ref{sec:main}.
	Then, we investigate the problem of estimating the secrecy capacity of the degraded wiretap channel characterized by the CMI.
	Finally the paper is concluded in Section~\ref{sec:conc}.

	\section{Preliminaries}\label{sec:pre}
	
\begin{figure}[t]
		\centering
		\tikzset{%
			block/.style    = {draw,fill=gray!10, thin, rectangle, minimum height = 2em,
				minimum width = 3em},
			sum/.style      = {draw, circle, node distance = 2cm}, 
			input/.style    = {coordinate}, 
			output/.style   = {coordinate} 
		}
		\resizebox{0.42\textwidth}{!}{
			\begin{tikzpicture}[thick]
			\tikzstyle{nod} = [draw=none,fill=none,right]
			\tikzstyle{add} = [draw,inner sep=0.1mm, circle]

			\draw 
			node at (0,0) [nod](M){$M$}
			node [block,right of=M,node distance=1.5cm](Enc){Encoder}
			node [block,right of=Enc,node distance=2.5cm](ch){$p(y|x)$}
			node [block,right of=ch,node distance=2.5cm](Dec){Decoder}
			node [block,below of=Dec,node distance=1.3cm](ch2){$p(z|y)$}
			node [nod,right of=ch2,node distance=1.3cm](Z){$Z^n$}
			[-latex] (5.5,0) |- (ch2)
			node [nod,right of=Dec,node distance=1.5cm](Mh){$\hat M$};
			
			\draw[-latex](M) --(Enc);
			\draw[-latex](ch2) -- (Z);
			\draw[-latex](Enc) --node[above]{$X^n$}(ch);
			\draw[-latex](ch) --node[above]{$Y^n$}(Dec);
			\draw[-latex](Dec) --(Mh);
			
			\end{tikzpicture}
		}
		\caption{Degraded wiretap channel.}
		\label{fig:DWTC}
	\end{figure}
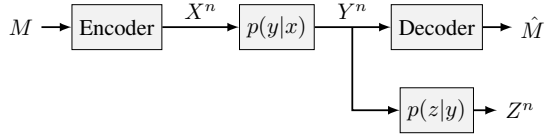

	For random variables $X, Y, Z$ jointly distributed according to $p(x,y,z)$, the CMI is defined as:
	\begin{align}
		I(X;Y|Z)=\KL{p(x,y,z)}{p(y,z)p(x|z)},
		\label{eq:CMI_def}
	\end{align}
	where $\KL{p}{q}$ is the KL-divergence between the probability density functions (\emph{pdf}s) $p(u)$ and $q(u)$, $u\in\mathbb{R}$,
	\begin{align}
		\KL{p}{q}=\int p(u)\log\frac{p(u)}{q(u)}du.
	\end{align}
	The CMI characterizes the capacity of communication systems such as channels with random states, network communication models like the relay channel or the degraded wiretap channel (DWTC)~\cite{el2011network}.
	For example, in the DWTC (see Fig.~\ref{fig:DWTC}), the secrecy capacity is:
	\begin{align}
		C_s=\max_{p(x)} I(X;Y|Z).\label{eq:DWTC}
	\end{align}
	
	
	%
	%
	%
	%
	%
	%
	
	By similar steps as in~\cite{barber2003algorithm}, we can bound \eqref{eq:CMI_def} from below:
	\begin{align}
		I(X;Y|Z)\geq E_{p(x,y,z)}\left[\log \frac{q(x|y,z)}{p(x|z)}\right],\label{eq:CMI_LB1}
	\end{align}
	where $q(x|y,z)$ is any arbitrary \emph{pdf}.
	The lower bound in~\eqref{eq:CMI_LB1} is tight if the conditional density functions $p(x|y,z)$ and $q(x|y,z)$ are equal.
	As discussed in~\cite{poole2018variational}, by choosing an energy-based density, one can obtain lower bounds for the CMI as stated in the following proposition.
	
	\begin{proposition}\label{pro:EB}
		For any function $f: \mathcal{X}\times \mathcal{Y}\times \mathcal{Z}\rightarrow\mathbb{R}$, let $M(y,z)=E_{p(x|z)}\big[ \exp f(x,y,z) \big]$, then the following bound holds  and is tight when $f(x,y,z)=\log p(y|x,z)+c(y,z)$:
		\begin{multline}
			I(X;Y|Z) \geq E_{p(x,y,z)}\left[f(x,y,z)\right]\\
			-E_{p(y,z)}\left[\log M(y,z)\right].
		\end{multline}
		\proof{
			Choosing $q(x|y,z)\coloneqq\frac{p(x|z)\,\exp f(x,y,z)}{M(y,z)}$ and substituting in~\eqref{eq:CMI_LB1} yields the desired bound.\qed
		}
	\end{proposition}	
	
	It is worth noting that computing the optimal $f(x,y,z)$ is non-trivial since we may not have access to the joint \emph{pdf}.
	Using Proposition~\ref{pro:EB} and Jensen's inequality, a variant of the Donsker--Varadhan bound~\cite{donsker1975asymptotic} can be obtained as follows:
	\begin{multline}\label{eq:LB_DV}
		I(X;Y|Z)\geq E_{p(x,y,z)}\left[f(x,y,z)\right]\\
		-\log E_{p(x|z)p(y,z)}\big[ \exp f(x,y,z) \big].
	\end{multline}
	Hereafter let the r.h.s.\ of~\eqref{eq:LB_DV} be denoted as $I_\DV$. The bound is tight when $f(x,y,z)=\log\frac{p(x|y,z)}{p(x|z)} + c$.
	
	In order to estimate $I_\DV$ from samples, an empirical average may be taken with respect to $p(x,y,z)$ and $p(x|z)p(y,z)$.
	Let $\mathcal{B}_\textnormal{joint}^b$ and $\mathcal{B}_\textnormal{prod}^b$ be a random batch of $b$ triples $(x,y,z)$ sampled respectively from $p(x,y,z)$ and $p(x|z)p(y,z)$; then the estimated $\hat I_\DV$ for an arbitrary choice of $f(x,y,z)$ is:
	\begin{align}\label{eq:LB_DV_est}
		\hat I_\DV^b &=
		\frac{1}{b}\sum\nolimits_{(x,y,z)\in\mathcal{B}_\textnormal{joint}^b} f(x,y,z) \nonumber\\
		&\quad -\log \frac{1}{b} \sum\nolimits_{(x,y,z)\in\mathcal{B}_\textnormal{prod}^b} \exp f(x,y,z) .
	\end{align}
	
	Since the outcomes of the previous estimator exhibit a relatively high variance (\cite{poole2018variational, mukherjee2019ccmi, mcallester2018formal}), the authors of~\cite{mukherjee2019ccmi} averaged the estimated results over several trials.
	However, due to the concavity of the logarithm in the second term of $I_\DV$, a Monte Carlo average of different instances of $\hat{I}_\DV$ results in an upper bound of $I_\DV$, while $I_\DV$ is itself a lower bound of the CMI.
	This issue can be resolved by taking a looser bound where the $\log$ term is linearized.
	A similar bound is obtained by Nguyen, Wainwright, and Jordan~\cite{nguyen2010estimating}, also adopted in~\cite{belghazi2018mine, mukherjee2019ccmi} to estimate the MI and which was denoted \textit{f-MINE} (since it corresponds to a variational representation for the f-divergence).
	The corresponding bound for the CMI is:
	\begin{multline}
		I(X;Y|Z)\geq E_{p(x,y,z)}\left[f(x,y,z)\right]\\
		- e^{-1}E_{p(x|z)p(y,z)}\big[ \exp f(x,y,z) \big], \label{eq:LB_NWJ}
	\end{multline}
	where we use $\log(u)\leq e^{-1}u$ in \eqref{eq:LB_DV}.
	In this paper, let $I_\NWJ$ refer to the r.h.s.\ of~\eqref{eq:LB_NWJ}.
	The bound is tight for $f(x,y,z)=1+\log\frac{p(x|y,z)}{p(x|z)}$, while 
	a Monte Carlo average of several instances of the estimator $\hat I_\NWJ$ is unbiased and, consequently, justified for estimating a lower bound on the CMI.
	
	Although the optimal choice for $f(x,y,z)$ is known, since the true joint density is unknown, it is non-trivial to compute. 
	Several approaches have been proposed to estimate $I_\NWJ$ for MI and CMI including~\cite{belghazi2018mine, poole2018variational} and \cite{mukherjee2019ccmi}.
	In~\cite{belghazi2018mine}, the searching space is restricted to functions generated via a neural network; then using minibatch gradient descent, the r.h.s.\ of~\eqref{eq:LB_NWJ} is optimized.
	Even though training the network can be unstable~\cite{poole2018variational}, their method is shown to scale better with dimension than the $k$-NN estimator for mutual information in~\cite{kraskov2004estimating}.
	Alternatively, \cite{poole2018variational} discusses estimators for log density ratio to be used in variational bounds, denoted there as Jensen--Shannon estimator.
	The motivation originates in the form of the functions $f(\cdot)$ that optimize these bounds.
	Similarly to estimate $\KL{p}{q}$, \cite{mukherjee2019ccmi} introduces a binary classifier using a neural network
	to discriminate samples from $p(x)$ and $q(x)$ and shows that by using cross-entropy loss, the density ratio present in the optimal $f(\cdot)$ can be estimated.
	The authors then applied this classifier to estimate the CMI using $I_\DV$.
	
	Despite the variational bound $I_\NWJ$ for the CMI resembling the one for the MI, in practice, obtaining the empirical estimation cannot be immediately extended due to the appearance of the conditional density $p(x|z)$ in preparing the batches.
	A proposed workaround in~\cite{mukherjee2019ccmi} is to compute the CMI as a difference between two MIs, i.e., $I(X;Y|Z)=I(X;Y,Z)-I(X;Z)$.
	However, given that both estimated MIs are obtained from lower bounds,
	it is not clear that their difference is a lower or upper bound of the desired CMI.
	
	Nonetheless, the authors of~\cite{mukherjee2019ccmi} do propose several methods to construct batches including GAN, variational autoencoder, and $k$-NN method.
	Then the classifier is trained using them to distinguish samples from either $p(x,y,z)$ or $p(x|z)p(y,z)$.
	In this work, we only focus on the $k$-NN method to create the batches and train the classifier similar to~\cite{mukherjee2019ccmi}.
	The main advantage of our work is to use $I_\NWJ$ to estimate the CMI, instead of $I_\DV$ as was applied in~\cite{mukherjee2019ccmi}, to resolve the Monte Carlo bias problem.
	This is crucial in the context of neural-based communication setups.
	The selection of a transmission rate above the true channel capacity, due to an overestimation induced by the bias, leads to a catastrophic increase in transmission errors.

	\section{Main results}\label{sec:main}
	
	In this section we explain the steps to estimate the CMI using neural networks.
	The estimator is subsequently exploited to compute the secrecy capacity of a degraded wiretap channel.
	
	\subsection{Constructing the batches}
	
	A major challenge in estimating the CMI for continuous random variables is dealing with the aforementioned conditional density function.
	Assume we have a dataset $\{(x_i,y_i,z_i)\}_{i=1}^n$ containing i.i.d.\ samples from $p(x,y,z)$.
	First, we split this dataset into train-set and test-set respectively to train the network and to calculate the estimate $\hat{I}_\NWJ$.
	The batch $\mathcal{B}_\textnormal{joint}^b$ is easily formed by taking $b$ random triples $(x_i,y_i,z_i)$ from the dataset.
	However, in order to construct $\mathcal{B}_\textnormal{prod}^b$, it is not clear how to take a sample $x$ according to $p(x|z)$ for a chosen pair $(y,z)$ distributed according to $p(y,z)$.
	
	Here we exploit the notion of $k$-NN as follows. 
	Consider $m:=b/k$ pairs $(y_l,z_l)$ chosen randomly from the dataset without replacement; they represent samples from $p(y,z)$.
	For each $z_l$, let $\mathcal{A}_{z_l}$ denote the set of indices corresponding to all its $k$ nearest neighbors (by Euclidean distance), i.e., $z_j$ is a $k$-NN of $z_l$ if $j\in\mathcal{A}_{z_l}$.
	Then, for all $m$ pairs $(y_l,z_l)$, we add to $\mathcal{B}_\textnormal{prod}^b$  all $k$ triples $(x_j,y_l,z_l)$ such that $j\in \mathcal{A}_{z_l}$.
	
	
	
	\subsection{Classifier}
	
	As previously seen, the optimal function $f(\cdot)$ in~\eqref{eq:LB_NWJ} is given by the log-likelihood ratio of two densities.
	In this work, we train a neural-based classifier to estimate that value.



	Consider a multi-layer neural network with the last layer implementing a sigmoid function mapping the output to a real value in $(0,1)$ denoted as $\omega$.
	The network is trained with the input data $\mathcal{B}_\textnormal{joint}^b$ and $\mathcal{B}_\textnormal{prod}^b$ namely joint and product class, where the objective is to classify the samples.
	For the training, we minimize the cross-entropy loss defined as
	\begin{align}
		L(\theta)=-\frac{1}{2b}\Bigl[&\sum\nolimits_{(x,y,z)\in \mathcal{B}_\textnormal{joint}^b}\log \omega(x,y,z) \nonumber\\
		&+ \sum\nolimits_{(x,y,z)\in \mathcal{B}_\textnormal{prod}^b}\log(1- \omega(x,y,z))\Bigr].
	\end{align}
	%
	After training, the classifier's prediction $\omega$ is an estimate of the probability of $(x,y,z)$ being produced from the joint density.
	%
	With a sufficient number of samples and by the central limit theorem, the loss function converges to its expected value.
	Thus, for a sufficiently trained network and large $b$, the output $\omega$ is close to the optimal $\omega^*$ which verifies that:
	\begin{align}
		\lambda(\omega^*) 
		\coloneqq \frac{\omega^*(x,y,z)}{1-\omega^*(x,y,z)}
		=\frac{p(x,y,z)}{p(x|z)p(y,z)}. \label{eq:optim_w}
	\end{align}
	In conclusion, by minimizing the cross-entropy loss, the trained network generates the output $\omega$ which determines the density ratio between the \emph{pdf}s  for a particular triple $(x,y,z)$.
	
	\subsection{Lower bound on $I(X;Y|Z)$}
	
	We choose to estimate the CMI with $I_\NWJ$ since it is unbiased and shown to be tight.
	We train a neural-network--based classifier as previously described and compute~\eqref{eq:optim_w}; this allows us to obtain the optimal $f^*(\cdot)$ for~\eqref{eq:LB_NWJ}, i.e., 
	\begin{align}
		f^*(x,y,z)=1+\log \lambda(\omega^*) .
	\end{align}
	Hence the empirical estimate is computed as:
	\begin{equation}\label{eq:est_I_NWJ_2}
		\hat I_\NWJ^b 
		= 1+\frac{1}{b}\sum\nolimits_{\mathcal{B}_\textnormal{joint}^b} \log \lambda(\omega^*)
		- \frac{1}{b} \sum\nolimits_{\mathcal{B}_\textnormal{prod}^b} \lambda(\omega^*).
	\end{equation}
	The steps of the estimation are stated in Algorithm~\ref{alg:bound}.
	
	\begin{algorithm}
		\SetAlgoLined
		\KwIn{$data=\{(x_i,y_i,z_i)\}_{i=1}^n$,trials $T$, batch size $b$}
		Split $data$ into $train\_set$ and $test\_set$\\	
		\For{t=1,\dots,T}
		{
			$\mathcal{B}^b_\textnormal{joint,train}, \mathcal{B}^b_\textnormal{prod,train}\gets$ Batch($\,train\_set\,,b$)\\
			Train the classifier with $\mathcal{B}^b_\textnormal{joint,train}, \mathcal{B}^b_\textnormal{prod,train}$\\
			$\mathcal{B}^b_\textnormal{joint,test}, \mathcal{B}^b_\textnormal{prod,test} \gets$ Batch($\,test\_set\,,b$)\\
			Compute $\hat I_\NWJ^{b,t}$~\eqref{eq:est_I_NWJ_2} using $\mathcal{B}^b_\textnormal{joint,test}, \mathcal{B}^b_\textnormal{prod,test}$
		}
		Monte Carlo average: $\hat I_\NWJ^{b}\gets \frac{1}{T}\sum_{t=1}^{T} \hat I_\NWJ^{b,t}$\\
		\Return $\hat I_\NWJ^{b}$
		\caption{Estimation of $I_\NWJ$}
		\label{alg:bound}
	\end{algorithm}

	\subsection{Numerical results}
	
	\begin{figure}[t]
		\centering
		\tikzset{%
			block/.style    = {draw,fill=blue!30, thick, rectangle, minimum height = 3em,
				minimum width = 3em},
			sum/.style      = {draw, circle, node distance = 2cm}, 
			input/.style    = {coordinate}, 
			output/.style   = {coordinate} 
		}
		\resizebox{0.5\columnwidth}{!}{
			\begin{tikzpicture}[thick]
			\tikzstyle{box}=[fill=blue!30,minimum width=1.5cm,minimum height=1cm,draw,rectangle]
			\tikzstyle{nod} = [draw=none,fill=none,right]
			\tikzstyle{add} = [draw,inner sep=0.1mm, circle]

			\draw 
			node at (0,0) [nod](X){$X$}		
			node at (1.5,0)[add,name=mult1]{\Large$+$}
			node [nod,above of=mult1,node distance=1cm](N1){$N_1$};
			\draw
			node [add,right of=mult1,node distance=1.7cm](mult2){\Large$+$}
			node [nod,above of=mult2,node distance=1cm](N2){$N_2$}
			node [nod,right of=mult2,node distance=1.2cm](Z){$Z$}
			;
			\draw[-latex](X) --(mult1);
			\draw[-latex](N1) --(mult1);
			\draw[-latex](N2) --(mult2);
			\draw[-latex](mult1) -- node[above]{$Y$}(mult2);
			\draw[-latex](mult2) --(Z);
			
			\end{tikzpicture}
		}
		\caption{DWTC with additive Gaussian noise.}
		\label{fig:DWTC2}
		\vspace{-2mm}
	\end{figure}
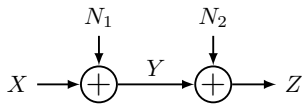

	\paragraph*{Wiretap channel}
	As motivated in Section~\ref{sec:intro}, estimators for CMI can be adopted to compute the capacity in communication systems and optimize encoders accordingly.
	One example is the DWTC (Fig.~\ref{fig:DWTC}) where a source is transmitting a message to a legitimate receiver while keeping it secret from an eavesdropper who has access to a degraded signal.
	The secrecy capacity~\eqref{eq:DWTC} can be estimated if samples of $(x,y,z)$ are available.
	Consider the channels $p(y|x)$ and $p(z|y)$ to have additive Gaussian noise with variance $\sigma^2_1$ and $\sigma^2_2$, respectively, then the model simplifies as shown in Fig.~\ref{fig:DWTC2}.
	So for any input $X$ such that $\textnormal{Var}[X]=P$, $I(X;Y|Z)$ can be computed as:
	\begin{equation}
		I(X;Y|Z)
		\leq \frac{1}{2} \log \bigg(1+ \frac{P}{\sigma_1^2}\bigg) -\frac{1}{2} \log \bigg(1+ \frac{P}{\sigma_1^2 + \sigma_2^2}\bigg),
	\end{equation}
	with equality when $X\sim\mathcal{N}(0,P)$.
	For our estimation, we consider $\sigma_1^2=1$ and the input $X$ to be zero-mean Gaussian with variance $P=100$; we collect $n=2e4$ samples of $(x, y,\allowbreak z)$ according to the described model and create batches of size $b=n/2$.
	The neural network in our experiment has two layers with 64 hidden ReLU activation units, and we use Adam optimizer with a learning rate $2e{-3}$ and 300 epochs. Final estimations are the averages of $T=20$ Monte Carlo trials.
	
\begin{figure}[t]
		\centering
		\includegraphics[width=.7\columnwidth]{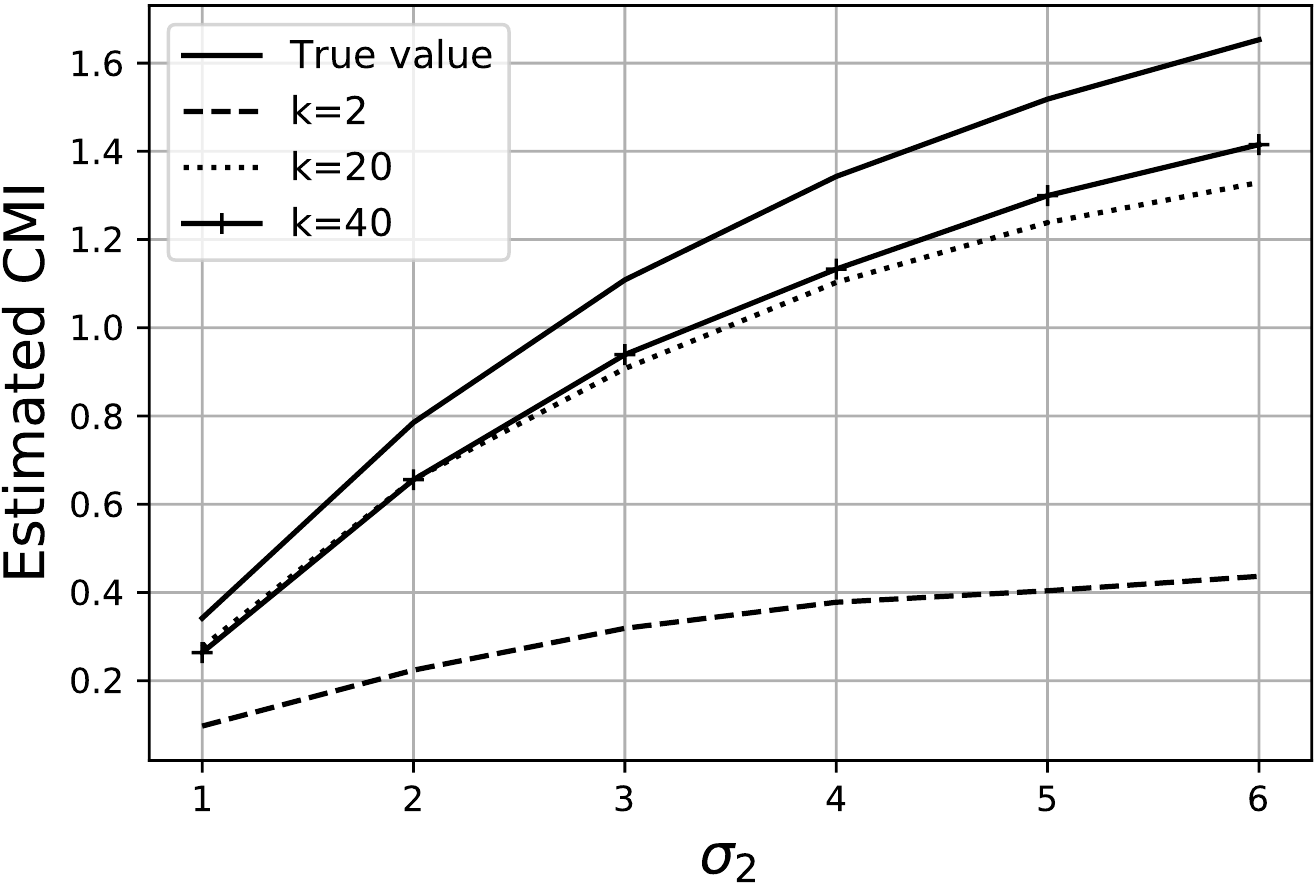}
		\caption{Estimated CMI vs. $\sigma_2$ for different choices of $k$.}
		\label{fig:result}
	\end{figure}

	The estimated CMI is depicted with respect to $\sigma_2$ in Fig.~\ref{fig:result} and for different choices of the number of neighbors $k$. For $\sigma_2=0$, i.e., the eavesdropper has access to the same signal as the legitimate receiver, the secrecy capacity is zero.
	It can be observed that increasing $k$ results in a better estimation.

\begin{figure}[t]
		\centering
		\includegraphics[width=.7\columnwidth]{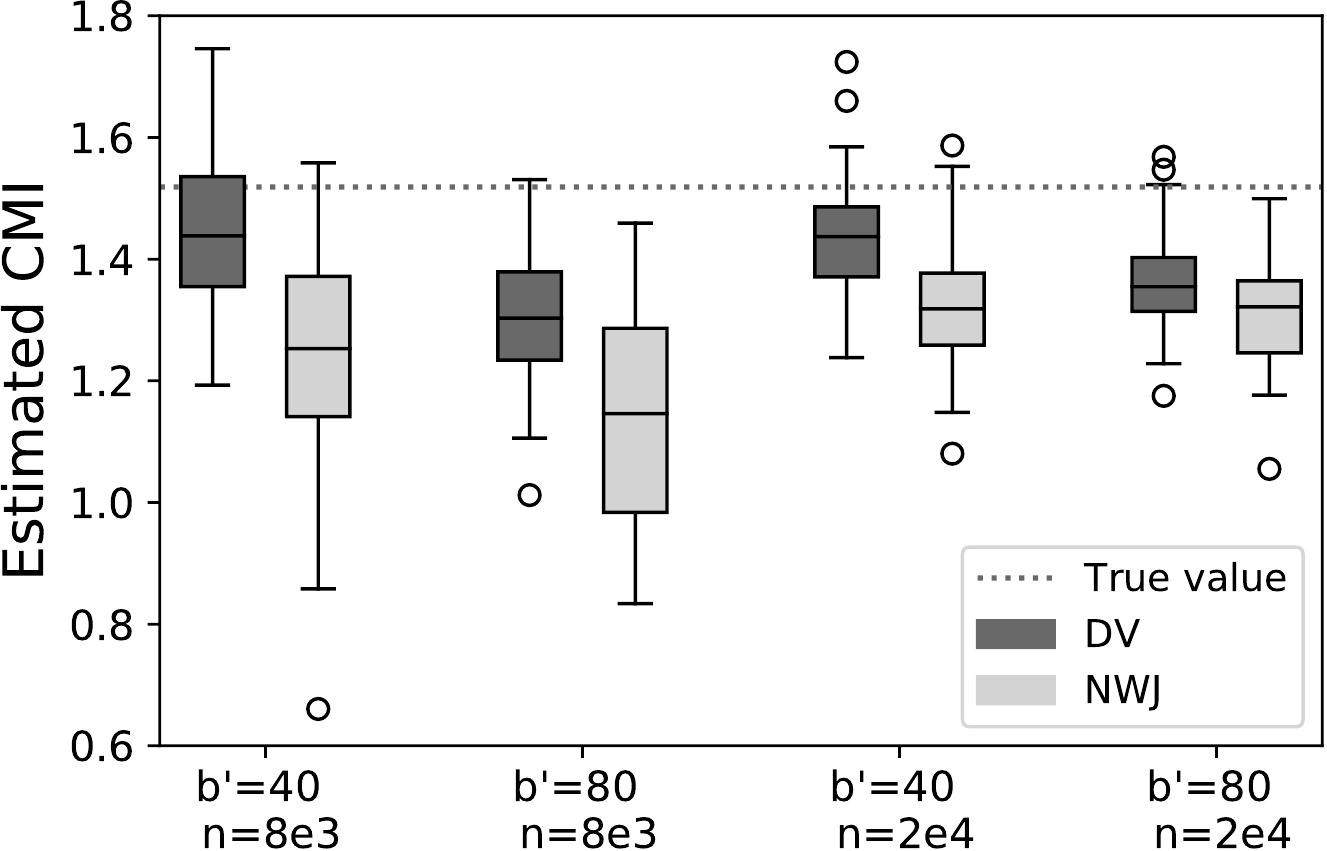}
		\caption{Effect of Monte Carlo average on $\hat I_\DV$ and $\hat I_\NWJ$.}
		\vspace{-1mm}
		\label{fig:result2}
		\vspace{-1mm}
	\end{figure}

	\vspace{-2mm}
	\paragraph*{Monte Carlo bias}
	To give some insights into the discussion on Section~\ref{sec:pre}, we compare the estimators $\hat I_\DV$ and $\hat I_\NWJ$ for the previously defined DWTC (fixing $\sigma_2=5$) in Fig.~\ref{fig:result2}.
	The classifier is trained with $b=n/2$ samples and choosing $k=40$; to compute the lower bound, the batches $\mathcal{B}^{\smash{b'}}_\textnormal{joint,test}$ and $\mathcal{B}^{\smash{b'}}_\textnormal{prod,test}$ are chosen with size $b'$ and the results are averaged over $T=20$ Monte Carlo trials.
	This procedure is repeated $50$ times with a new dataset each time to obtain the boxplots.

	It can be observed that, for small values of $b'$, averaging the lower bound~\eqref{eq:LB_DV_est} over trials is more likely to yield an overestimation of the true CMI compared to~\eqref{eq:est_I_NWJ_2}.
	This is a joint effect of the Monte Carlo bias and the small sample size --the latter affecting both estimators in terms of variance.
	This justifies our decision in using $\hat I_\NWJ$.
	While larger $n$ and $b'$ can significantly decrease the variance of the estimation, the bias can be reduced by increasing $k$ as shown in Fig.~\ref{fig:result}.

	\section{Conclusion}\label{sec:conc}
	
	In this paper, we investigated the problem of estimating the conditional mutual information using a neural network.
	This was motivated by its application in learning encoders in communication systems.
	Since the conventional methods to estimate information-theoretic quantities do not scale well with dimension, recent works have proposed to estimate them utilizing neural networks.
	
	Challenges of the extensions from estimators of mutual information have been discussed.
	Additionally, we argued on the advantages of our method in terms of estimation bias and showed the performance in estimating the secrecy transmission rate in a degraded wiretap channel.
	As a future direction, this method can be applied to other communication schemes and coupled with an optimizer for encoders.

	{\small
		\bibliographystyle{IEEEtran}
		\bibliography{IEEEabrv,ref}
	}

\end{document}